\documentclass[onecolumn, superscriptaddress, prmaterials, amsmath,amssymb,noeprint,12pt]{revtex4-2}

\usepackage{hyperref}
\hypersetup{backref,colorlinks=true,allcolors=prussianblue}

\usepackage{graphicx}
\usepackage[caption=false]{subfig}

\usepackage[table]{xcolor}
\definecolor{prussianblue}{rgb}{0.0, 0.19, 0.33}

\usepackage{setspace}
\usepackage{float}
\usepackage{indentfirst}

\usepackage{tcolorbox}
\usepackage{todonotes}

\def\eq#1{Eq.~\eqref{eq:#1}}

\UseRawInputEncoding

\begin{document}

\author{Mauricio Ponga}
\author{Mohamed Hendy}
\author{Okan K. Orhan}
\address{Department of Mechanical Engineering, University of British Columbia, 2054 - 6250 Applied Science Lane, Vancouver, BC, V6T 1Z4, Canada}

\author{Swarnava Ghosh} 
\address{National Center for Computational Sciences, Oak Ridge National Laboratory, Oak Ridge, TN 37830, USA}

\title{Effects of the local chemical environment on vacancy diffusion in multi-principal element alloys}

\begin{abstract}
Multi-principal element alloys (MPEAs) are exciting systems showing remarkable properties compared to conventional materials due to their exceedingly large compositional space and spatially varying chemical environment. However, predicting fundamental properties from the local chemical environment is challenging due to the large scale of the problem. To investigate this fundamental problem, we employ a combination of atomistic simulations (using \emph{ab-initio} and molecular dynamics) and convolutional neural networks (CNNs) to evaluate point defect and migration energies in an equimolar CoFeCrNi MPEA. We show how energies of point defects can be predicted with reasonable accuracy using a small subset of local chemical environments. Using the CNNs, we develop a lattice Monte Carlo simulation that computes the migration path and diffusivities of vacancies. Remarkably, our work illustrates how the local chemical environment leads rise to a distribution function of the point defect energies, which is responsible for the macroscopic diffusivity of vacancies. In particular, we observed that vacancies get trapped in super basins surrounded by large migration and connected with low migration energy states. As a result, vacancy diffusivity is highly dependent on the environment and could change several orders of magnitude for a given temperature. Our works illustrate the importance of understanding properties in MPEAs depending on the local chemical environment and the ability of CNN to provide a model for computing energies in high-dimensional spaces, which can be used to scale things up to higher-order models.  \footnote{This manuscript has been authored in part by UT-Battelle, LLC, under contract DE-AC05- 00OR22725 with the US Department of Energy (DOE). The publisher acknowledges the US government license to provide public access under the DOE Public Access Plan (http://energy.gov/downloads/doe-public-access-plan).}
\end{abstract}

\maketitle

\section{Introduction}
Multi-principal element alloys (MPEAs), also known as high entropy alloys, combine multiple elements at similar atomic molar proportions ~\cite{doi:10.1002/adem.200300567,CANTOR2004213,Yeh2006,Gild2016,Oses2020}. This alloying strategy has been attracting increasing attention since these materials often exhibit remarkable properties, such as high strength \cite{CHENG20113185,CAI2019281}, high ductility \cite{LI201835} at cryogenic and high temperatures \cite{XIAN2017229}, high toughness, and corrosion resistance~\cite{GALI201374,CHEN201639}. The large compositional space of MPEAs offers limitless possibilities to design new materials with exciting properties for industrial applications, especially under extreme environments \cite{Li2022}. As a result, a vast spectrum of experimental works have been published using conventional and more advanced manufacturing techniques \cite{ROJAS2022}, showing a broad range of properties.

In MPEA, the \emph{local chemical environment} -the spatial and chemical distribution of elements in a given region of the alloy- is a highly variable quantity that could include random solid solution, but, in some cases, short-range ordering within the material~\cite{BALDERESCHI197599,PhysRevB.27.2587,MAURIZIO2003178,MIRACLE2017448,SRO1, SRO2, SRO3}. This local chemical complexity leads to a series of quantum mechanical interactions between atoms in the alloy, leading to forces that will produce local distortions in the atoms. %Even though the community refers to these interactions as \emph{elastic distortions}, these result from complex quantum interactions that grant MPEAs their extraordinary properties. 
Understanding the effects of these complex interactions is the key to MPEAs design. 

Even though MPEAs' remarkable properties are often attributed to their composition and microstructure  \cite{Tsai2013}, there still is a lack of understanding of the fundamental mechanisms that lead to these properties. Computational work has been primarily focused on developing strategies for alloy design using a myriad of techniques, including empirical design parameters, thermodynamics methods~\cite{ZHONG2019108060}, machine learning (ML)~\cite{Kaufmann2020}, and multi-objective optimization techniques~\cite{10.3389/fmats.2021.816610}. On the other hand, mechanistic models to describe mechanical behavior have been limited to the understanding of pinning of dislocations with composition \cite{MARESCA2020,MARESCA2020b}, and prediction of defect energy based on composition \cite{Aidhy:2021} and novel mechanisms of deformation depending on the environment \cite{Cao2020}. This limitation is partially due to the fact that the complex local chemical environment in MPEAs makes these properties spatially dependent on the material, including an additional layer of complexity in modeling such problems.

Diffusion in MPEAs has been a heated subject in the scientific community early since the inception of the sluggish diffusion term was coined \cite{Ye2018,Daw2021,Dabrowa2020}. The first quantitative work produced by Tsai and co-workers provided relevant experimental data that set the following works on MPEAs diffusion \cite{Tsai2013}. Other experimental results with radioactive isotopes have confirmed these findings for bulk \cite{Vaidya2017} and grain boundaries studies \cite{Vaidya2016}. A number of previous work have also reported the chemical role in the activation energies \cite{Dabrowa2016,Thomas2020,KOTTKE2020236}, while others have also pointed out the importance of including atomistic details in the migration of vacancies, which remains unavailable in coarse-graining models \cite{Wang2022,Roy2022}. 

In this work, we propose a combination of atomistic simulations, ML, and Monte Carlo methods to investigate vacancy diffusion in MPEAs. We first compute with \emph{ab-initio} and classical molecular static  simulations vacancy formation and migration energies in equimolar CoFeCrNi and show the dependency on the chemical environment of these values. The topology and chemical information of each defect are then used to train two CNNs to predict the vacancy formation and migration (forward and backward) energies. By doing so, we develop a model where the local chemical environment is used to compute point quantities in the alloy. The CNNs are then integrated into a Monte Carlo algorithm to simulate vacancy migration in equimolar CoFeCrNi at several temperatures.

Our results show that vacancies can find migration paths with low energy barriers but eventually get trapped in basins delimited by higher energy barriers. This effect is responsible for the diffusivity of vacancy from a macroscopic viewpoint. As a result, the vacancy diffusivity can vary several orders of magnitude for a given vacancy in a specific local chemical environment. Besides providing a fundamental understanding of vacancy diffusion in MPEAs with the corresponding dataset \cite{Ponga:2022VacML}, our work introduces a framework to combine data-driven approaches based on atomistic simulations with well-established models in materials science to simulate relevant macroscopic properties in these materials. The framework can then be modified for other phenomena by recurring to mechanistic models while retaining crucial chemical information about these defects. 

\section{Methodology} \label{Section:Methodology}

Let us now define the problem at hand that we want to explore here. Let $N_e \in \mathbb{N}$ be the number of atomic species (or elements) and $N_s \in \mathbb{N}$ be the number of sites in a multi-principal element metallic alloy. At the same time, $ \mathbf{x} \in \mathbb{R}^{N_e}, ~ \mathbf{x} = \bigg(\frac{N_1}{N_s}, ~\frac{N_2}{N_s}, ~\frac{N_i}{N_s}, \ldots, \frac{N_{N_e}}{N_s} \bigg)$ denotes the vector of global atomic molar fractions representing the alloy composition and $\mathbf{x}_i$ be the $i-$th component of that vector. $N_{i} \in \mathbb{N}$ is the number of atoms of the $i-$th element with the constrain $\sum_i^{N_e} N_i = N_s$. For an equiatomic alloy, as considered here, $N_i = N_a ~\forall ~i$, and $N_s = N_e N_a$. Next, suppose that the alloy is a \emph{random solid solution} (RSS) therefore, the possible set of spatial configurations is a vast number. These random configurations are represented by the array of atomic positions together with the chemical elements ${\bf q}_j = \{q_1^j,~q_2^j,~q_3^j,m~^j \}$, with ${\{ {\bf q}_j \} \in \mathbb{R}^{4\times N_s}} $ where ${\bf q}_j$ is a vector containing the position in the reference configuration and the chemical species $m=1,2,\ldots,N_e$. Theoretically, a random species can occupy each atomic site in the set $N_e$. 

We want to investigate the formation energy of point defects, including vacancies ($f^{v}$) and self-interstitial atoms ($f^{s}$) in multi-principal element metallic alloys. However, from a computational perspective MPEAs represent a challenge as the number of configurations or representations of the same system is exceedingly large. A calculation provided in the Appendix~\ref{Section:ComputationalGeneration} shows these vast number of configurations. To circumvent this issue, we use the following approach. For a given random site $i$, we define the \emph{local chemical environment} as the array of atomic positions and species surrounding this site in a small cluster containing $N_c$ atoms and represented by ${\{ {\bf q}_i\} \in \mathbb{R}^{4\times N_c}}$ as shown in Fig.~\ref{fig:Schematic}. Our goal is to obtain the approximated formation values using the local chemical environment and a surrogate model, i.e., convolutional neural network. 

Face-centered cubic single crystals with a equimolar CoFeCrNi were generated. Thousand of RSS composition were obtained using a homogenous random generator in Matlab while imposing equimolar composition. Several samples were generated and the statistics of the generator were characterized by investigating the \emph{local chemical environment} of each cell. To this end, we computed for each atom the atomic molar fraction of each element as a function of the different neighbor shells (up to the the four nearest neighbor shell (4NN)), i.e., 

\begin{equation}
\mathbf{c}_i = \sum_j^{N_c} \frac{1}{N_c} \begin{pmatrix} n_a^1,~n_a^2,\ldots,~n_a^{N_e}    \end{pmatrix},
\end{equation}
where $\mathbf{c}_i$ is a vector of length $N_e$, $n_a^i$ are the number of atoms of the $i-$the species, and $N_c$ are the atoms in the selected neighbor shell in the pristine lattice. The {\bf Supplementary Materials (SM)}~\cite{SeeSM} Figs.~\ref{fig:performance} and \ref{fig:chem_env} show the performance and statistics of the random cell generation, respectively.

\subsection{Atomistic simulations}
\emph{Ab-initio} simulations were performed within the Kohn-Sham density functional theory \cite{PhysRev.136.B864,PhysRev.140.A1133}. We used the Quantum Espresso software \cite{Giannozzi_2009} using a supercell containing $N_s = 256$ atoms.  A plane-wave kinetic energy cutoff of 150 Rydberg was employed  with the PBE functional \cite{PhysRevB.33.8822,PhysRevB.46.6671}. The equilibrium lattice constant was found to be $3.506$ \AA ~when the pristine sample was minimized at zero pressure. For all calculations, we relax the cell and the atom positions such that the system is stress free and the atomic forces are below  $10^{-3}$ Ry$\cdot$Bohr$^{-1}$.
The formation energy $f^{v}$ was computed using \eq{Formation} the energies of the pristine sample ($E_0(N_s)$), and the energy of the sample when one atom was removed ($E_d(N_s-1)$) and the chemical potential of the atomic species of the vacancy ($\mu_d$) \cite{PhysRevMaterials.2.013602}.

\begin{equation} \label{eq:Formation}
f^{v} = E_d(N_s-1) - E_0(N_s) + \mu_d.
\end{equation}
To compute the chemical potential of each species, we performed the procedure detailed proposed in an earlier work \cite{PhysRevMaterials.2.013602} and briefly summarized in the {\bf SM}~\cite{SeeSM}. All atoms in the sample were removed one at the time to compute the $N_s$ formation energies.

Classical molecular statics simulations were also used to compute the formation energy of the defects. The Large-scale Atomic/Molecular Massively Parallel Simulator (LAMMPS) \cite{Plimpton:1995} package was used to compute all molecular statics (MS) simulations with the embedded atom model (EAM) potential developed by Farkas and Caro ~\cite{Farkas:2018}. Once the random samples were generated, simulation cells for vacancy and interstitial calculations were performed by removing/including atoms of several species. The nudge elastic band (NEB) method was used to compute the migration paths and energy values \cite{henkelman2000climbing,henkelman2000improved}. First, samples were relaxed with forces below $10^{-6}$ eV$\cdot$\AA$^{-1}$ and zero pressure. Then, the initial and final stages of the vacancies were generated and eight replicas were used to interpolate the intermediate steps. A spring constant of $1$ eV$\cdot$\AA$^{-2}$ was used in all simulations. A random perturbation of 0.1 \AA~was imposed to the atoms' position before the minimization was performed with the quickmin algorithm. 

\begin{figure}[t]
\centering
\includegraphics[width=0.45\textwidth]{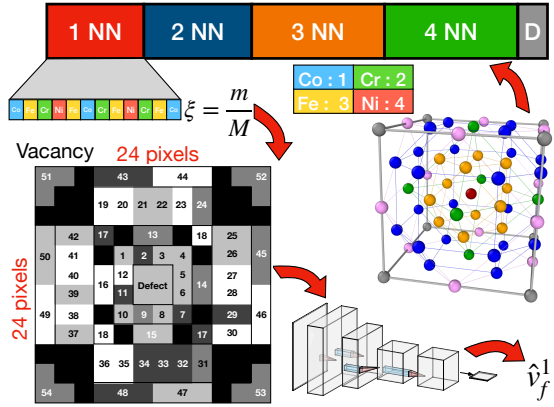}
\caption{Pixel map generation for CNN pattern recognition and data regression. Neighbor atoms types around the point defect are scanned and arranged in a 2D array for each simulation. For each atom type an integer $m \in [1,M]$ is assigned. Then, the pixel map is obtained as an array of $24\times24$ where the pixel intensity is $\xi = m/M$ as shown in the schematic. }
\label{fig:Schematic}
\end{figure}

The neighbor atoms surrounding the defects were recorded along with the formation energies. Fig.~\ref{fig:Schematic} shows the pixel map generation scheme. Up to four neighbor shells (4NN) were used in our calculations (involving up to 55 sites) and arranged in a 1D array. The atomic species $m=1, \ldots, M$ were then dumped for each defect in a row format, and a pixel of intensity $\xi = m/M$ was generated for each simulation. Several arrangements were tested, and after a trial and error procedure, we arrived at a simple but effective way of arranging this information in the pixel map. Ultimately, the images were characterized by a 576-pixel map arranged in a $24\times24$ input array. Examples of different vacancy sites are shown in Fig.~\ref{fig:Schematic}. Large dark areas at the edge of the figures are unfilled pixel cells with intensity zero.

\subsection{Convolutional neural network architecture}
After generating the pixel maps, we developed a CNN as shown in Fig.~\ref{fig:CNN_architecture}. The CNN architecture involved a $24\times24 \times 1$ input layer, three deep neural layers with a convolution, normalization, activation layer (ReLU), and averaging pooling were included. The first layer included eight filters ($24\times24\times16$), followed by thirty-two ($12\times12\times32$) and sixty-four ($6\times6\times64$) filters in the second and third deep layers, respectively \cite{Gonzalez2007,Mahendran2017}. The regression layer included a deep layer with a convolution operation, a normalization, and ReLU layers, followed by a dropout and fully connected layer with one neuron to predict the formation energy. The same architecture was used to predict the forward, and backward migration energies, except two bit maps of  $24\times24 \times 2$  (with the vacancy in the initial and final positions) were used as input, and two neurons were used in the last layer for forward and backward energies. A schematic view of the CNN architecture is shown in Fig.~\ref{fig:CNN_architecture}. The convolution layers had a filter with dimensions $2\times2$ with a padding option set to {\texttt{same}, whereas the average pooling layer had a stride of $2\times 2$. The dropout probability was set to 0.10. In total, about 29500 learnable values are used to train the model. 

The weights and biases were obtained by training the network with the data obtained and post-processed in pixel maps as described above. The following options were used for the optimization of the weights and biases using the ADAM solver with a squared \texttt{ GradientDecayFactor} of 0.99 \cite{Kingma2015}. A batch size of 32 with a maximum of 45 epochs was used. The initial learning rate was set to 0.005 for all cases with a piece-wise learning rate schedule and drop factor of 0.1 and a period of 20. For training, validation, and testing purposes, the data set was split into three groups, using 70\% of the data set for training, 15\% for validation, and 15\% for testing. Vacancies, self interstitial atoms (SIAs), and forward and backward energies were trained separately for convenience. The root mean square error (RMSE) and the R$^2$ coefficient are used as performance metrics of the CNNs.

\begin{figure}[t]
\centering
\includegraphics[width=0.45\textwidth]{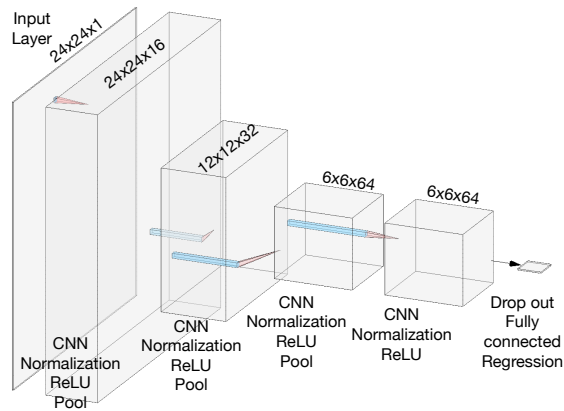}
\caption{CNN architecture used to recognize the point defect pattern structure and prediction of formation values. Three CNN layers are used for pattern recognition, while the last CNN is used for regression of the formation, migration, forward and backward transition values. }
\label{fig:CNN_architecture}
\end{figure}

\subsection{Vacancy diffusion using ML and MC}
Next, we propose to study the diffusion of vacancies in MPEAs using the trained CNN within the Monte Carlo method. For shortness, we will refer to this method as MLMC. To this end, we used the surrogated model to investigate vacancy formation and migration energies. This information can be used in an MC method that exchanges vacancies between nearest neighbor sites in an FCC lattice \cite{Andersen2019,Kimari2020,KOTTKE2020236}. We now assume that vacancy hops are not correlated in time (e.g., $f^v_i(t) \neq f^v_i(\{f^v_j(t)\}$), and their trajectory is solely determined by the energy landscape given by the changing local chemical environment.

Since the formation and migration energies vary depending on the composition, a kinetic law was required. Consider that a vacancy is at the $i-$th site of the lattice and could migrate to another $j-$site (state). The rate constant for this migration is $k_{ij}=\nu_0\exp(\frac{-Q_m^{i\rightarrow j}}{k_BT})$, where $\nu_0$ is the attempt frequency. Here, $\nu_0$ was estimated using the average value of the quasi-harmonic stiffness matrix for each atom at a given temperature $T$ \cite{Ariza:2012,Ponga2017a,Ponga2018c,Mendez:2021}. However, recent work has pointed out that $\nu_0$ could drastically change for MPEAs \cite{Prefactor:2022}. A histogram with the harmonic frequencies computed with this framework can be seen in Fig.~\ref{fig:VibFreq} in the {\bf SM}~\cite{SeeSM}. $Q_m^{i\rightarrow j} = f^v_j + e_{i\rightarrow j}$ is the activation energy, usually defined as the sum of the formation energy ($f^v_j$) and the migration energy ($e_{i\rightarrow j}$). Since each vacancy have a finite number of sites to migrate ($p=12$), the total scape rate can be described as $k_t = \sum_j^{p} k_{ij}$ \cite{Andersen2019}. 

The following procedure was employed to select the escape state from the $i-$th site. The vacancy formation energies of all possible nearest sites ($f^v_j$) were computed using the surrogate model. Then, a probability that the vacancy will migrate to a $j-$site was calculated using the following expression
\begin{align} 
\rho_{i\rightarrow j} = \begin{cases}
1 & \text{ if } f_i^v \geq f_j^v \\
\exp(-(f^v_j-f^v_i)/k_\mathrm{B} T)& \text{ if } f_i^v < f_j^v.
\end{cases}
\end{align} 
The site with the highest probability was selected. If several sites had the same probability, then the one with lower $f_j^v$ was selected. To include stochasticity in the escape trajectory, we performed an additional step if the $f_i^v \geq f_j^v$. A randomly drawn number, $\rho_1$, was selected. Then, $\rho_1$ was compared to a threshold $\epsilon = 0.05$, such that if $\rho_1 < \epsilon$, a random nearest site was selected instead of the $j-$site. This modification was done to consider different scenarios where thermal vibrations can push the system out of equilibrium and the results were robust when this condition was not included. 

Since the vacancy has a decaying probability function for the actual time of escape, the total time that the system is advanced is $\Delta t = \ln(1/\rho_2)/k_{t}$, where $\rho_2 \in [0,1]$ was another randomly drawn number.  The vacancies' time evolution can now be modeled, provided that a (surrogate) model is available to compute these energies as a function of the local chemical environment. The CNN introduced earlier was used here.

Once the model had been calibrated, we proceeded to run several simulations for randomly selected vacancies using 500,000 sites equimolar CoFeCrNi. Several temperatures of interest were set, from 300 to 1500 K at increments of 150 K. All simulations were run for a very long time, more than 100 ns, and out of the range of traditional MD simulations.

\section{Results}

 \begin{figure}
\centering
\includegraphics[width=0.45\textwidth]{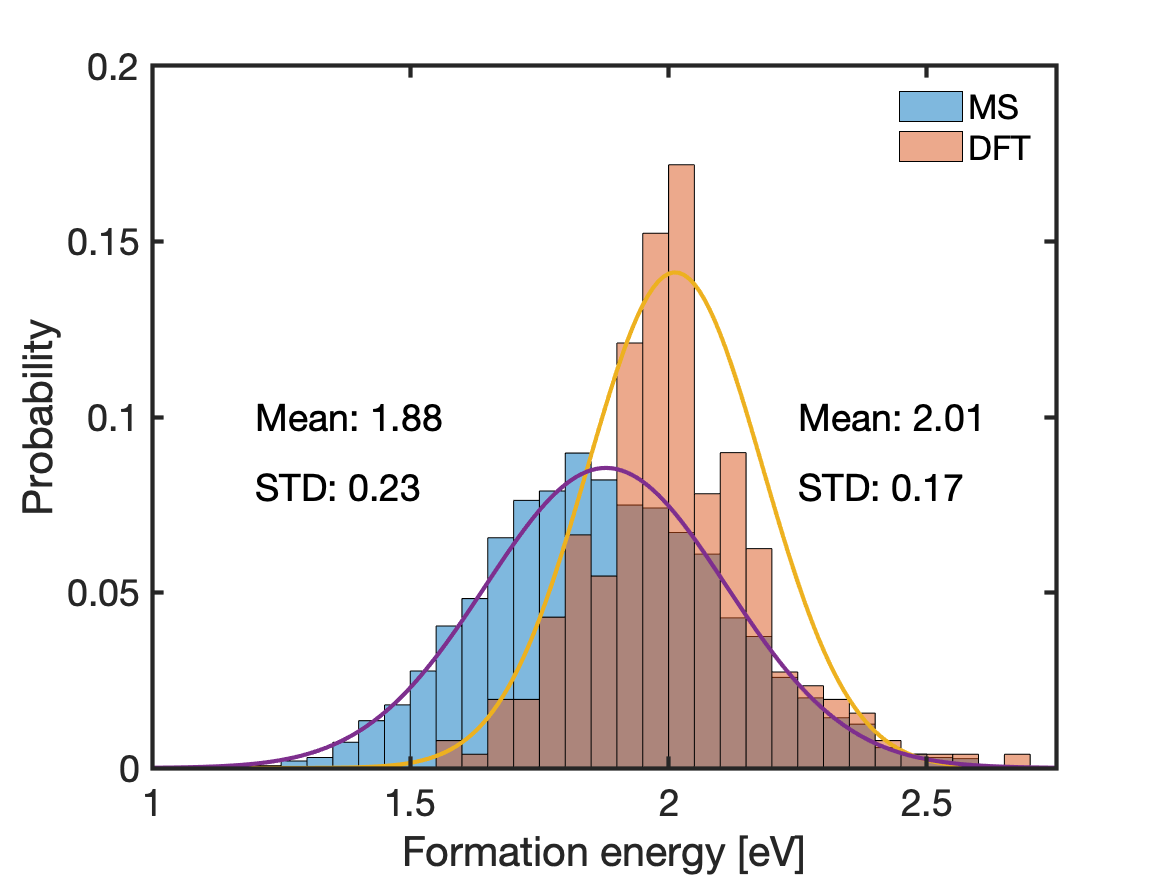}
\caption{Vacancy formation energy ($f^{v}$) histogram as obtained with DFT and MS. A  Gaussian distribution is shown for both techniques shown their respective mean and standard deviations. Distributions were obtained for an equimolar CoFeCrNi sample containing 256 atoms in DFT and 500,000 atoms in MS. Mean and standard deviation (STD) values are shown in eV.}
\label{fig:VacHist}
\end{figure}

\subsection{Point defect energy calculations}

Next, we compare the formation values for vacancies between DFT and MS. Our goal is to provide a qualitative and quantitative analysis between the two techniques and point out their similarities and differences. To this end, we computed the formation energy for 256 vacancies using the same computational cell for DFT, and several thousand of formation energies with MS. The $f^v$ energy distributions obtained with DFT and MS are shown in Fig.~\ref{fig:VacHist}. As can be seen from the histograms, the $f^v$ has a normal distribution independently of the technique used. Remarkably, DFT values have a mean of $\sim2$ eV with a standard deviation of $0.17$ eV, whereas MS has a mean of $\sim1.88$ eV with a standard deviation of $0.23$ eV. %The distribution of values arose from the individual local interactions between vacancies and their local chemical neighborhoods.

 A clear difference between the DFT and MS values was observed, pointing to MS's limitations in predicting complex interactions near the vacancies. Moreover, the error difference between DFT and MS for individual vacancies does not correlate well with any function, pointing to complex chemical and quantum interactions near the vacancy that cannot be easily described. Despite these differences between DFT and MS, we point out that the local chemical environment is represented in the MS simulations, which can be used to feed the CNN model. 

 \subsection{Migration energies}

 Next, we proceed to investigate the statistics of the forward and backward vacancy migration energies. As shown in Fig.~\ref{fig:ForwHist}, the forward and backward migration energies follow a Gaussian distribution, centered around 1.15 eV. The standard deviations of the forward and backward migration energies were $\sim0.166$ and $\sim0.171$ eV, respectively. Figs.~\ref{fig:TransEnergy} shows the behavior of the transition energy for a vacancy in the FeCoCrNi HEA as a function of the reaction coordinate. The analysis of the transition energy performed over $\sim6000$ vacancies illustrates the chemical complexity of the HEA. On average, the transition energy has a zero net change (same initial and final energy) as indicated by the blue line and a maximum barrier value of $\sim1$ eV. However, individual vacancies will migrate to different energy levels that could be either lower or higher energy states, as indicated by the light blue shaded area that shows the standard deviation of the transition energy ($e_{i\rightarrow j}$) as a function of the reaction coordinate. The standard deviation is about $\sim0.25$ eV for the forward and backward energies. 

\begin{figure}[t]
\centering
\includegraphics[width=0.45\textwidth]{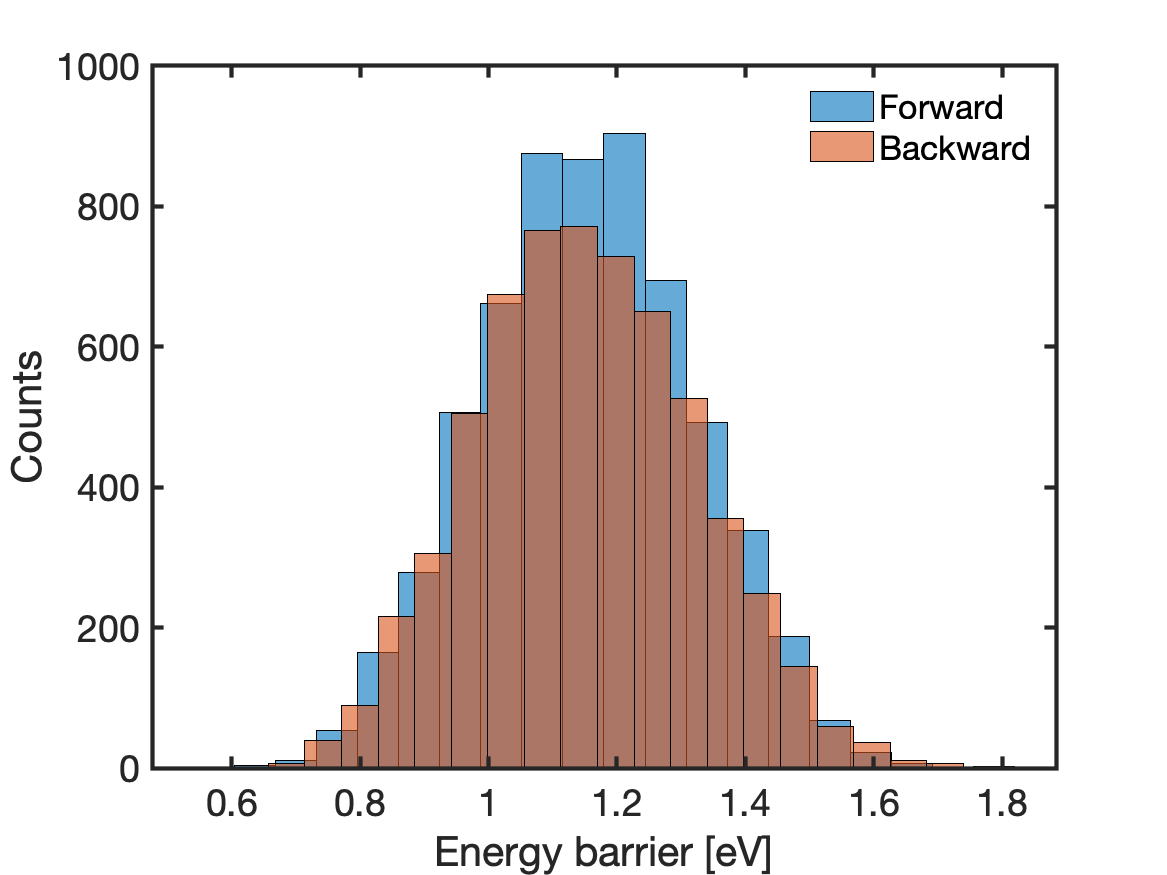}
\caption{Forward and backward migration energy histograms obtained with NEB simulations. The average value was 1.15 eV for forward and backward migration energies. The standard deviations were $\sim0.166$ and $\sim0.171$ eV for forward and backward migration, respectively.}
\label{fig:ForwHist}
\end{figure}

Remarkably, the maximum and minimum paths also show a large spread (as high as 1.75 eV and as low as 0.55 eV) of the transition energy, as indicated by the solid red lines. While \emph{a priori} these values might seem too small to elucidate differences, we should remind that vacancies will tend to diffuse depending on their transition energy which factors into an Arrhenius law $\propto \exp(\frac{-Q_m}{k_BT})$. Thus, vacancies will migrate at different rates as $T$ increases, inevitably leading to complex non-linear diffusion.

\begin{figure}[t]
\centering
\includegraphics[width=0.45\textwidth]{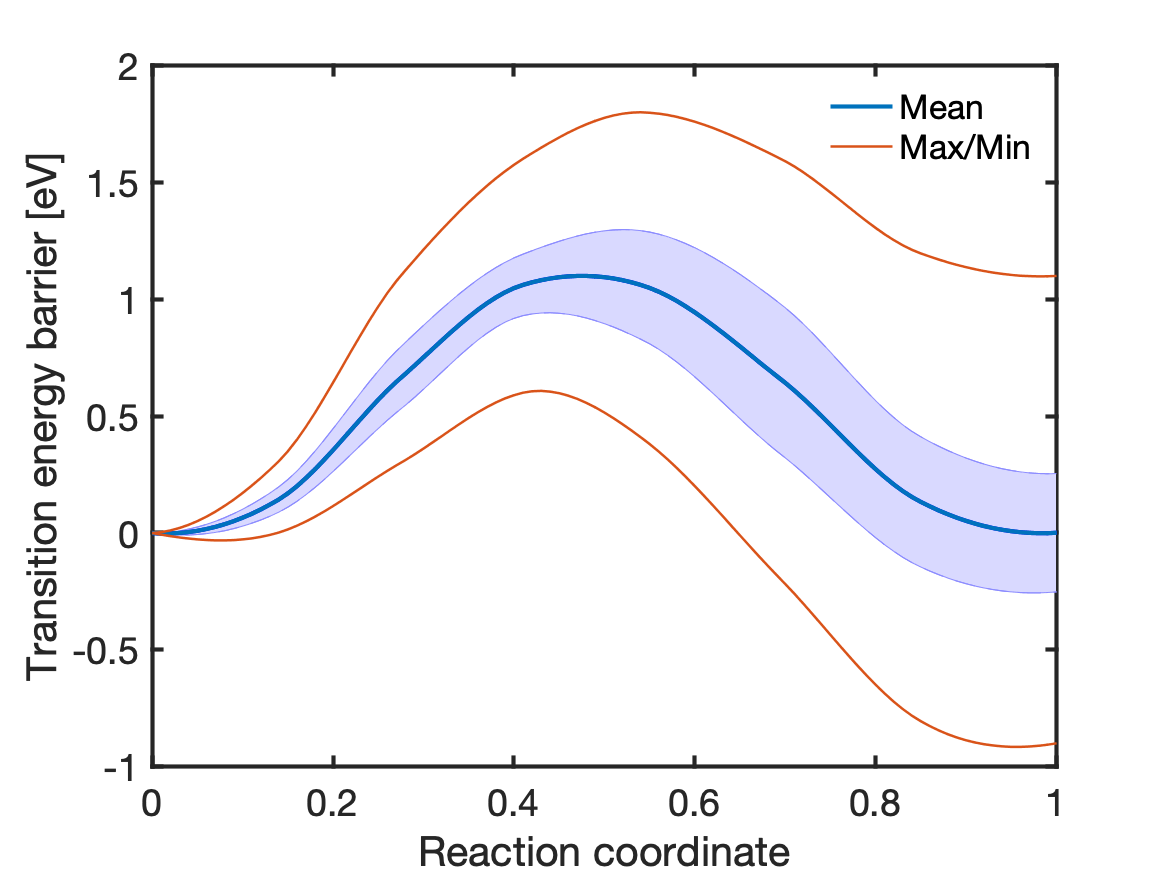}
\caption{Transition energy values for equimolar  CoFeCrNi high-entropy alloy as a function of the reaction coordinate. Solid blue line indicates the average value of the transition energy, whereas the blue shaded area indicates the standard deviation of all ($\sim6000$) considered atomistic simulations used in this work. Solid red lines indicate the maximum and minimum values around the transition path.}
\label{fig:TransEnergy}
\end{figure}

\subsection{Machine learned formation and migration energies}
Next, we show the performance of the CNN with the previously described data set and optimization parameters. All values are dimensionalized using the maximum ($f_{\text{max}}^v$) and minimum ($f_{\text{min}}^v$) formation values. 
The dimensionalized values were computed as $\overline{f}^v = f^v - f_{\text{min}}^v/ \Delta f^v$, with $\Delta f^v = f_{\text{max}}^v- f_{\text{min}}^v$.
Fig.~\ref{fig:VacAll} shows the evolution of the predicted \emph{vs.} true values of the formation energy ($\overline{f}^{v}$) for the test set. The central dashed line represents a perfect prediction, whereas the other two dashed lines show the region where 95\% of data points are located. The RMSE was $\sim0.05$ with a $R^2 = 0.91$.

Similarly, Fig.~\ref{fig:SIAAll} shows the evolution of the predicted \emph{vs.} true values of the self interstitial atom formation energy ($\overline{f}^{s}$) for the validation set. The RMSE for the model was $\sim0.05$ with a $R^2 = 0.96$. We also see that due to the significantly larger difference between values, three different regions of data points are observed.  

The transition energy values were processed and fed into a CNN described in the Methodology section. 
The results of the CNN for the forward and backward transition energies are shown in Fig.~\ref{fig:Forw}. Noteworthy, these plots show only the testing data that was not used for training or validation. While greater spread was observed compared to the formation energies, the CNN reasonably predicted the transition energies with RMSE values of 0.07 and R$^2$ of 0.81.

\begin{figure}[t]
\centering
\includegraphics[width=0.45\textwidth]{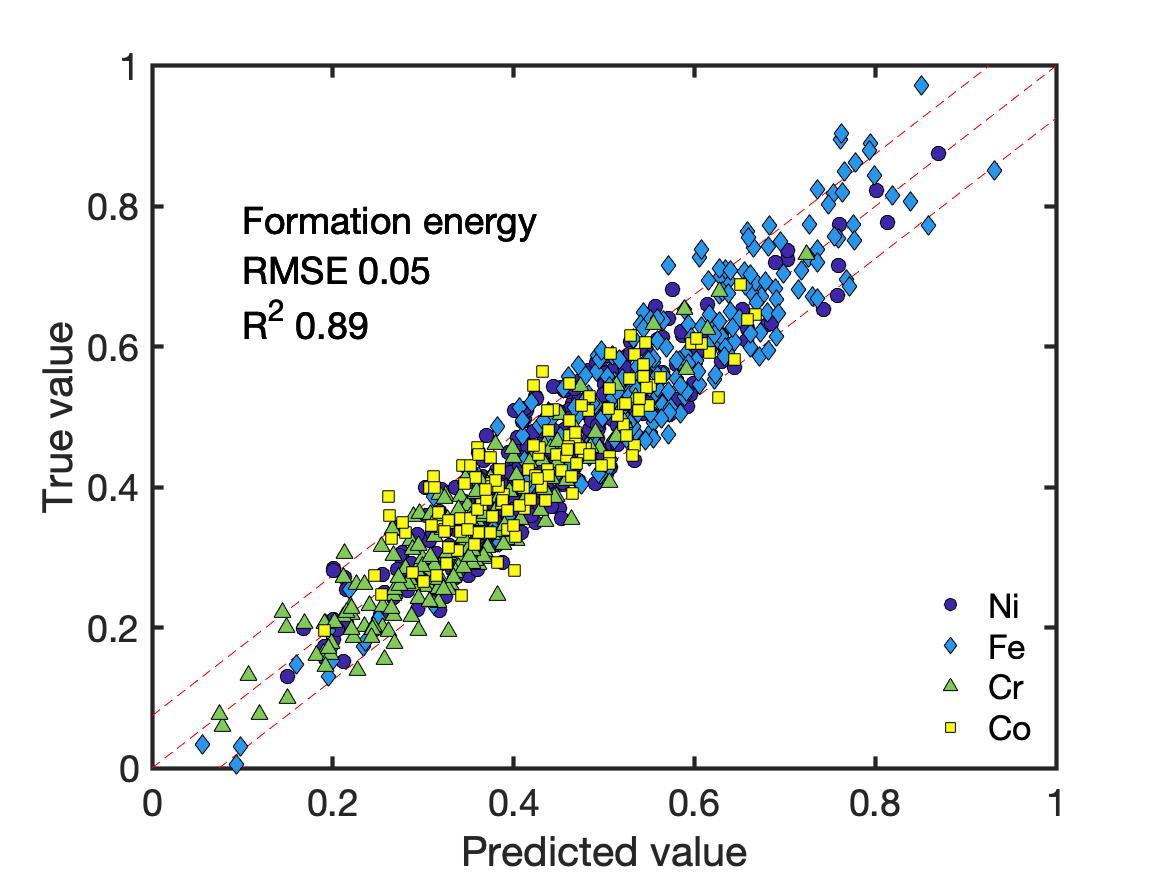}
\caption{Predicted \emph{vs.} true values of vacancy formation energies for a CoFeCrNi high-entropy alloy as predicted with the trained CNN using the test set.}
\label{fig:VacAll}
\end{figure}

\begin{figure}[t]
\centering
\includegraphics[width=0.45\textwidth]{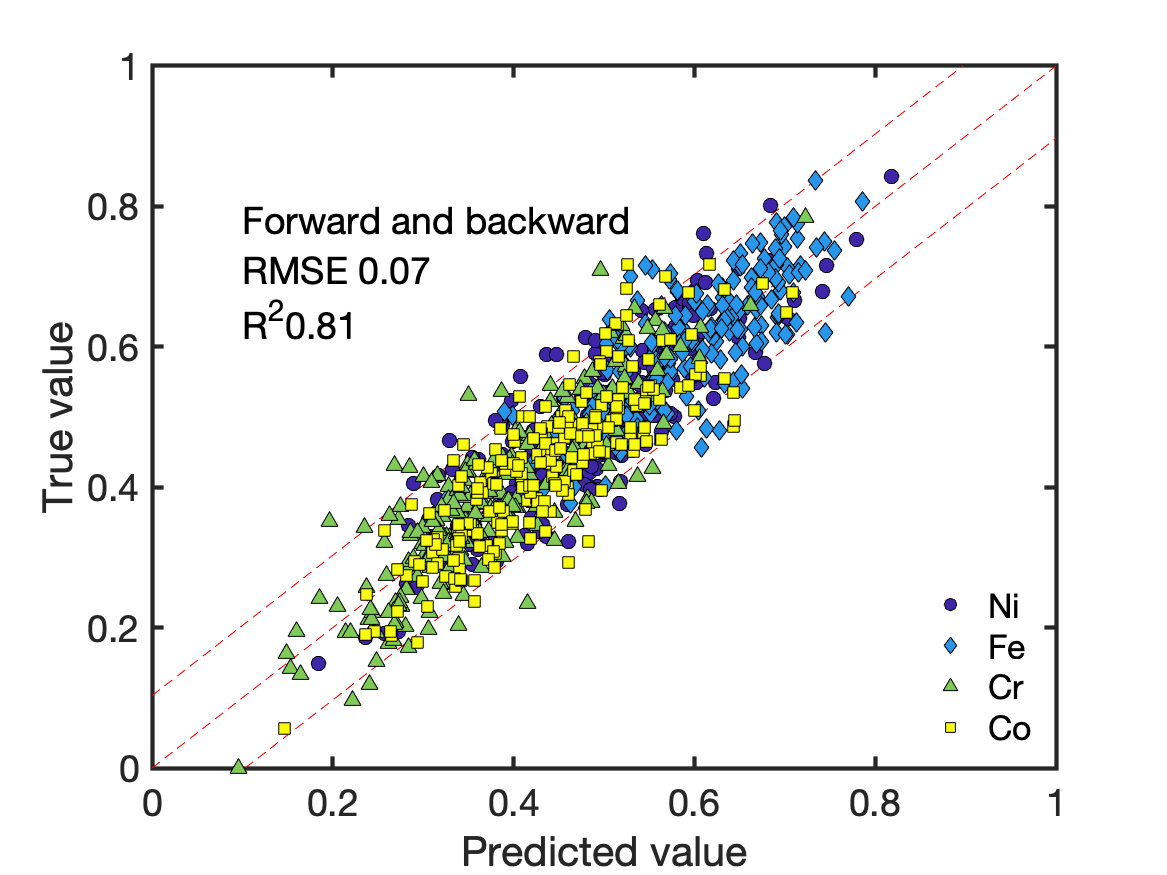}
\caption{Predicted \emph{vs.} true values of forward and backward migration energies for a CoFeCrNi high-entropy alloy as predicted with the trained CNN using the test set. The dashed lines indicate the exact (middle) and one standard deviation from the exact value. Points with different colors indicate different types of elements removed to generate the vacancy. }
\label{fig:Forw}
\end{figure}

\subsection{Vacancy diffusivity using the MLMC method}

Having developed a surrogate model for prediction of vacancy formation and migration energies, we proceeded to use this information into a Monte Carlo method. Remarkable, the use of the CNN allow us to retain local chemical environments as in atomistic simulations while the MC method alloys to simulate transition paths more easily than in MD. Fig.~\ref{fig:Diffusivity} shows the vacancy diffusivity as a function of the temperature from 450~K to 1500~K at intervals of 150~K obtained with the Monte Carlo method using the CNN models. Due to the  large widespread values of the vacancies formation and migration energies, the diffusivity shows an extensive range of values that can oscillate several orders of magnitude for a given temperature.  
Due to smaller formation values for Cr and Co, vacancies move faster in the lattice compared to Ni and Fe on \emph{average}, although some latter element vacancies can move significantly faster than the median values and vice versa. Experimental measures of the diffusivity in CoFeCrNi MPEA obtained by  \cite{Tsai2013,Vaidya2017} are shown with $\diamond$~symbols, which seem to overlap well with the median value of the distribution obtained.

\begin{figure}[t]
\centering
\includegraphics[width=0.45\textwidth]{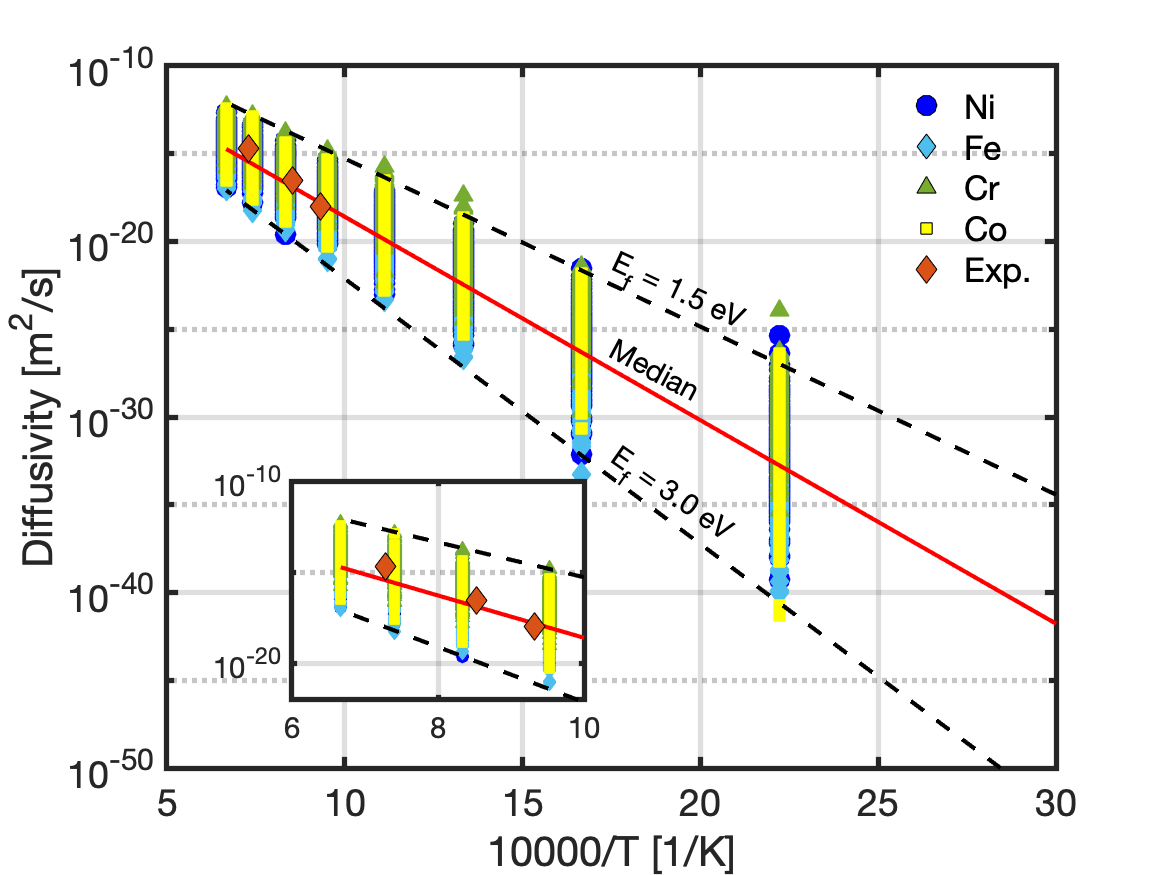}
\caption{Diffusivity of vacancies as a function of the temperature from 450 K to 1500 K at intervals of 150 K in a CoFeCrNi high-entropy alloy simulated with the MLMC method. Solid line indicates the median value, while the two dashed lines indicate the bounding values with activation energies of 1.5 and 3.0 eV, respectively. Experimental values shown with $\diamond$ obtained from \cite{Tsai2013,Vaidya2017}. The inset shows a close up view of the high-temperature range.}
\label{fig:Diffusivity}
\end{figure}

The MLMC simulations demonstrate a complex diffusion behavior that is dominated by the local chemical environment that is revealed thanks to the CNN model for vacancy formation and migration. The simulations also show that due to the large widespread value of the diffusivity, the median value is more appropriate than the average value. 

\section{Discussions}
Atomistic simulations carried out with \emph{ab-initio} simulations and interatomic potentials within MS have shown that vacancy formation and migration values are highly dependent on the local chemical environment. This energy formation variation means that \emph{properties of individual vacancies are heavily influenced by the local chemical environment}; in line with several modeling works published before \cite{PhysRevMaterials.2.013602}. This dependency on the formation and migration values leads to highly non-linear interactions between defects that depend on the local chemical environment. Unfortunately, the dependence on fine details near defects poses challenges to modeling other derivative properties, such as vacancy diffusivity in MPEAs \cite{Vaidya2016,Vaidya2017}. 

To remedy this issue, we showed how data-driven techniques could be used to develop surrogated models with reasonable accuracy using relatively small data sets. Considering the ample configurational space of MPEAs (see the Methodology section), the CNNs predicted formation and migration values with remarkable accuracy, reaching an RMSE of 5 and 7 \%, respectively (see Figs.~\ref{fig:VacAll}~and~\ref{fig:Forw}). The use of CNNs retained atomic level information to compute the formation and migration energies, which would have been extremely difficult to model using mechanistic models \cite{Thomas2020,Tsai2013}. Thus, CNN and similar data-driven approaches could be beneficial in predicting energy values of defects in MPEAs with atomic level information as input. Noteworthy, the evaluation of the networks is computationally inexpensive and much smaller than running a full scale atomistic simulation. Therefore, CNN and akin models can be handy for linking atomistic information to mechanistic models, as illustrated with the MLMC technique. 

The introduction of the CNNs into the MC method allowed us to simulate the diffusion of vacancies at different temperatures for large time scales compared to regular MD simulations \cite{Wang2022} while retaining the local chemical environment information. The MC simulations showed that, on average, vacancies follow the average activation energy trend; however, some vacancies can significantly deviate from this behavior. Fig.~\ref{fig:distElements} shows the probability density function of the computed diffusivities for each individual element at $T=1200$ K using a semilogarithmic plot. As seen in the graphical representation, elements with much higher activation energy could have specific vacancies moving either faster or slower than the expected behavior (e.g., Ni and Fe vacancies). In addition, due to the ample range of activation values, the median or mid-value are more relevant than the average property value. This observation suggests that several realizations of the same MPEA might be required in properties with significant widespread to obtain meaningful predictions. The same behavior was found when the temperature was changed. Fig.~\ref{fig:distTemp} shows the probability density function of all vacancies in an equimolar CoFeCrNi MPEA at temperatures between 900 to 1500 K. Remarkably, the range of diffusivities increased with temperature. Conversely, as temperature decreased, the ratio between slowest and fastest vacancies increased. 

\begin{figure}[t]
\centering
\includegraphics[width=0.45\textwidth]{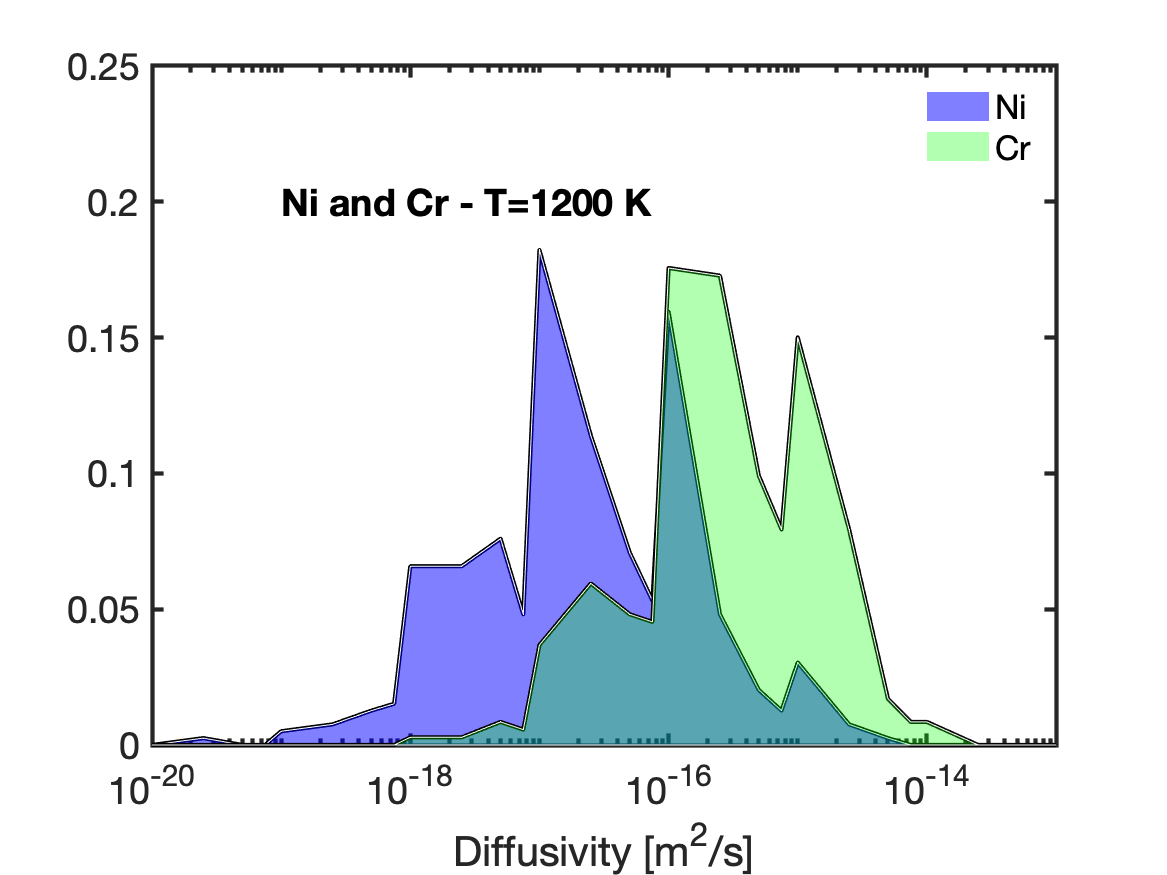}
\includegraphics[width=0.45\textwidth]{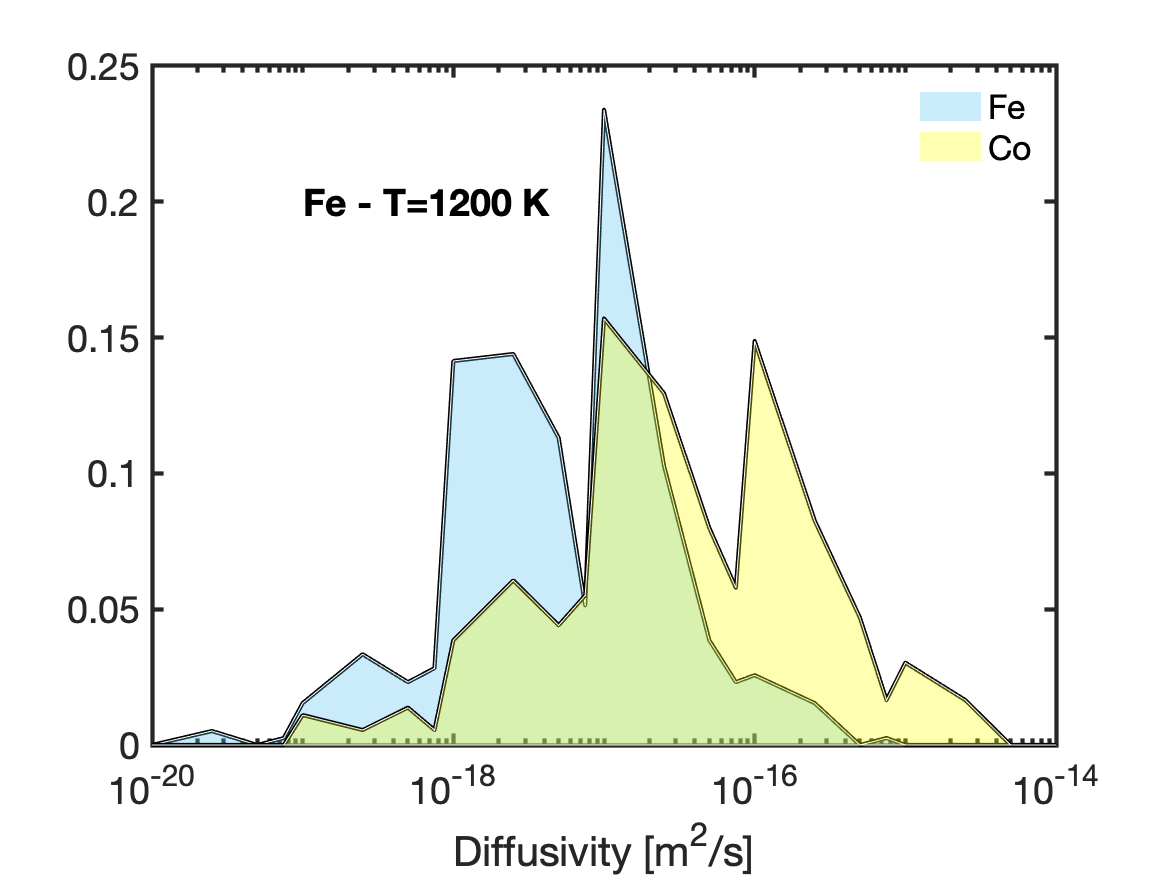}
\caption{Probability distribution for vacancy diffusivities for each individual element in an equimolar CoFeCrNi MPEA computed with the MLMC method at $T=1200$~K. }
\label{fig:distElements}
\end{figure}

Vacancy migration in MPEAs have been a subject of high controversy since the introduction of sluggish diffusion \cite{Dabrowa2020,jien2006recent,Kottke2020,Ye2018,Daw2021}. A key difference between MPEAs and other traditional alloys is that the effect of the local chemical environment is more evident in all sites of the material, while in traditional alloys, the chemistry plays a much more important role in defects such as grain/twin boundaries and dislocations. For instance, MC simulations of diffusion in a bulk material (e.g., Ni) would have required only one formation and migration energy values, whereas in MPEAs required much more. This contrast between MPEAs and traditional alloys makes predictions in the former group more subtle, and several realizations of the same system have to be taken into consideration. 

Remarkably, in MPEAs, vacancies move in the lattice until they find transition paths that require high energy levels, according with the hypothesis used to formulate sluggish diffusion. However, it was also observed that vacancies can find migration paths that require much lower migration energies than the average value as pointed in Fig.~\ref{fig:TransEnergy}. Therefore, diffusion of vacancies can be either (much) slower or faster than the median values as shown in Figs.~\ref{fig:Diffusivity}~and~\ref{fig:distElements}. These lower energy transition paths seem to have been missed in previous discussion of MPEAs and play an important role in the diffusivity of vacancies. Noteworthy, we found that when all paths are considered over a large number of vacancies and realizations, the median values of the diffusivity are in remarkable agreement with experimental measures obtained with high temperature experiments \cite{Tsai2013,Vaidya2017}. Therefore, our work provides a fundamental understanding of vacancy diffusion in MPEA, reconciling opposite views proposed in MPEAs.

When calibrated to an Arrhenius law, all simulated vacancies  ($\sim6000$) using different random generations seem to be bounded between two activation values of 1.5 and 2.5~eV, with $D_0^u = 10^{-6}$ and $D_0^l = 10^{-8}$~m$^2\cdot$s$^{-1}$ for the upper and lower bounds, respectively. This feature can be used in mechanistic models at the continuum scales where individual neighborhoods and local chemical environments cannot or might not be practically possible to consider.

\section{Conclusion}
To sum up, we have performed a series of atomistic simulations with \emph{ab-initio} and molecular statics to compute the formation and migration energies of point defects in equimolar CoFeCrNi. We have shown that these values depend on the individual local chemical environment and this fact can be used to train data-driven models such as CNNs to predict such values with remarkable accuracy. We have shown that these surrogate models can then be used in higher scales models, retaining all the atomic information. We have shown how vacancies can diffuse over the lattice and, eventually, get trapped in super basin bounded with large activation energy values getting trapped between specific regions of the sample. On the other hand, we also observed that due to the large number of migration paths, vacancies can migrate using paths that are much lower energy than the average values. This dependency on the atomistic environments results in a large widespread of diffusivity values for vacancies of the same element. Our work provides an avenue to retain valuable atomic information to describe defects in MPEAs, and to use this information in higher scale formulations by recurring to data-driven models. This can therefore be of remarkable importance in mechanistic models to understand the mechanics of materials. 

Finally, while the current CNNs are trained at the reference state -without any deformation-, the models can be modified with the aid of continuum theories to include the change of the formation and migration energy due to different deformation gradients and activation volumes. This can be useful to reduce the amount of data needed to calibrate the ML model, which can be then used in continuum models as shown in this work. Of remarkable interest, high temperatures creep behavior driven by diffusion of vacancies towards dislocation cores can be an exciting avenue to explore using the proposed models with such modifications. 

\section{Acknowledgment}
We acknowledge the support of New Frontiers in Research Fund (NFRFE-2019-01095) and from the Natural Sciences and Engineering Research Council of Canada (NSERC) through the Discovery Grant under Award Application Number 2016-06114. M.H. gratefully acknowledges the financial support from the Department of Mechanical Engineering at UBC through the Four Years Fellowship. %
This research was supported in part through computational resources and services provided by Advanced Research Computing at the University of British Columbia. This research used resources of the Oak Ridge Leadership Computing Facility, which is supported by the Office of Science of the U.S. Department of Energy under Contract No. DE-AC05-00OR22725.

\section{Data availability}
All code used in this manuscript and the associated data can be found in the  \href{https://github.com/Mponga/HEA_vacancy_mcml.git}{online site}.

\section{Author contributions}
M.H., O.K.O and M.P. carried out the MS simulations. M.P carried out the data analysis, and developed the ML and MLMC models. S.G. and O.K.O. performed the \emph{ab-initio} simulations. M.P. wrote the paper with inputs from all the other authors. 

\begin{appendix}
\section{Computational cells generation} \label{Section:ComputationalGeneration}

From a computational perspective, it is interesting to determine the number of possible configurations or realization available to simulate MPEAs within the context mentioned in the Methodology. To account for the number of possible representations, we assume that atoms of the same species are identical, and the permutation of these atoms does not generate a different configuration. Taking the first species $N_a^{1}$, we want to know how many possible combinations we can arrange these atoms into the $N_s$ sites. Using the combination notation, we find the number of possible combinations for element one given by Eq.~\ref{Eq:Combination1}.

\begin{equation} \label{Eq:Combination1}
C_1(N_s,N_1) = \begin{pmatrix}
N_s \\ N_1 
\end{pmatrix} =\frac{N_s!}{N_1! (N_s-N_1)!},
\end{equation}
where the $!$ symbol denotes factorial operation.

Next, having eliminated $N_1$ sites from the originally available $N_s$, we are left with the \emph{remaining} $R_1=N_s-N_a^1$ available sites. If we now want to arrange the $N_2$ atoms belonging to the second element, the number of combination possibles are then  

\begin{equation}
C_2(N_s-N_1,N_2) = \begin{pmatrix}
R_1 \\ N_2
\end{pmatrix} =\frac{(N_s-N_1)!}{N_2! (N_s-N_1-N_2)!}.
\end{equation}

It is now easy to see that the total number of representations of the alloy is then, 

\begin{align} 
N_\mathrm{cells} & =   \begin{pmatrix}
N_s \\ N_a^1
\end{pmatrix}
\begin{pmatrix}
N_s-N_1 \\ N_2
\end{pmatrix}\ldots
\begin{pmatrix}
N_s-N_{N_e} \\ N_{N_e} 
\end{pmatrix} \nonumber \\ 
& =   \prod_{i=1}^{N_e} \begin{pmatrix} R_i \\ N_i \end{pmatrix} ,
\end{align} 
where $R_i=N_s- \displaystyle \sum_{j=1}^{i-1} N_j$.
\begin{align} 
R_i = \begin{cases}
N_s & \text{ if } i=1 \\
N_s- \displaystyle \sum_{j=1}^{i-1} N_j & \text{ if } i \neq 1
\end{cases}
\end{align} 

Noteworthy, this is an extremely large number even for small $N_s$. 
For instance, taking $N_s=100$ and $N_a^i =25$ (corresponding to an equiatomic four elements HEA),  we obtain $N_\mathrm{cells}\sim1.61 \times 10^{57}$. For a surface calculation with $N_s=45$, it leads to $ N_\mathrm{cells}\sim1.90\times10^{28}$ for a five elements equimolar HEA. This simple calculation illustrates the need for developing coarse-grained models for HEAs.

\end{appendix}

\clearpage
\onecolumngrid
\raggedbottom
%%%%%%%%%% Prefix a "S" to all equations, figures, tables and reset the counter %%%%%%%%%%
\setcounter{section}{0}
\setcounter{equation}{0}
\setcounter{figure}{0}
\setcounter{table}{0}
\setcounter{page}{1}
\makeatletter
\renewcommand{\theequation}{S\arabic{equation}}
\renewcommand{\thefigure}{S\arabic{figure}}
\renewcommand{\thetable}{S\arabic{table}}
\renewcommand{\thepage}{SM\arabic{page}}

\renewcommand{\bibnumfmt}[1]{[S#1]}
\renewcommand{\citenumfont}[1]{S#1}

\begin{center}
{\Large \bf Supplementary Materials}
\end{center}

\section{Alloy generation}
We developed the following \emph{ad-hoc} criterion to assess the different random generations. Knowing the target composition ($\mathbf{x}_i$), one could analyze what samples were closer to this goal with the smallest standard deviation while at the same time imposing the smallest range of composition (maximum and minimum composition for each element in the shell). Then, simple performance metrics can be defined as,

\begin{equation} \label{eq:criterion}
\gamma = \sum_i^{N_s} (\mathbf{c}_i-\mathbf{x}_i)^2 \mathbf{\sigma}_i^2 (\mathbf{max}_{\mathbf{c}_i}-\mathbf{x}_i)^2 (\mathbf{min}_{\mathbf{c}_i}-\mathbf{x}_i)^2.
\end{equation}

The significance of \eq{criterion} is that if each site on the generated sample hits the target composition \emph{locally}, then $\gamma$ is zero. Higher values of $\gamma$ indicate large differences in the local environment compared to the target composition. Noteworthy, while the criterion evaluates the performance of the random generation, it should not be interpreted as an indicator for \emph{superior} computational generation.

We briefly analyze the random solid solution generation algorithm using the analysis tools described above.  Fig.~\ref{fig:performance} presents the performance parameter $\gamma$ for thousand random generations cells containing ~256 atoms. It is interesting to see that a few generations have remarkably low $\gamma$ which indicates closer compositions to the target. This is illustrated in Fig.~\ref{fig:chem_env}(left), where the atomic molar fraction as a function of the neighbor shells (NS) is shown. Noteworthy, the standard deviation of this generation is large (about 0.12 to 0.05) indicating local variability in the composition of the alloy. Interestingly, the analysis of the performance parameter indicates that the most common samples are not correlated to lower values of $\gamma$, but with values that oscillate between $10^{-4}- 10^{-3}$ Fig.~\ref{fig:chem_env}(center). This suggest that i) \emph{the most probable RSS generations} are in this range; and ii) even when homogeneous random generator are used some short-range order (SRO) appear in the microstructure. Generations with $\gamma \ge 10^{-3}$ are rare and could have large differences in compositions up to the third neighbor shell (3NN), e.g., Fig.~\ref{fig:chem_env}(right).

\begin{figure}[t]
\centering
\includegraphics[width=0.45\textwidth]{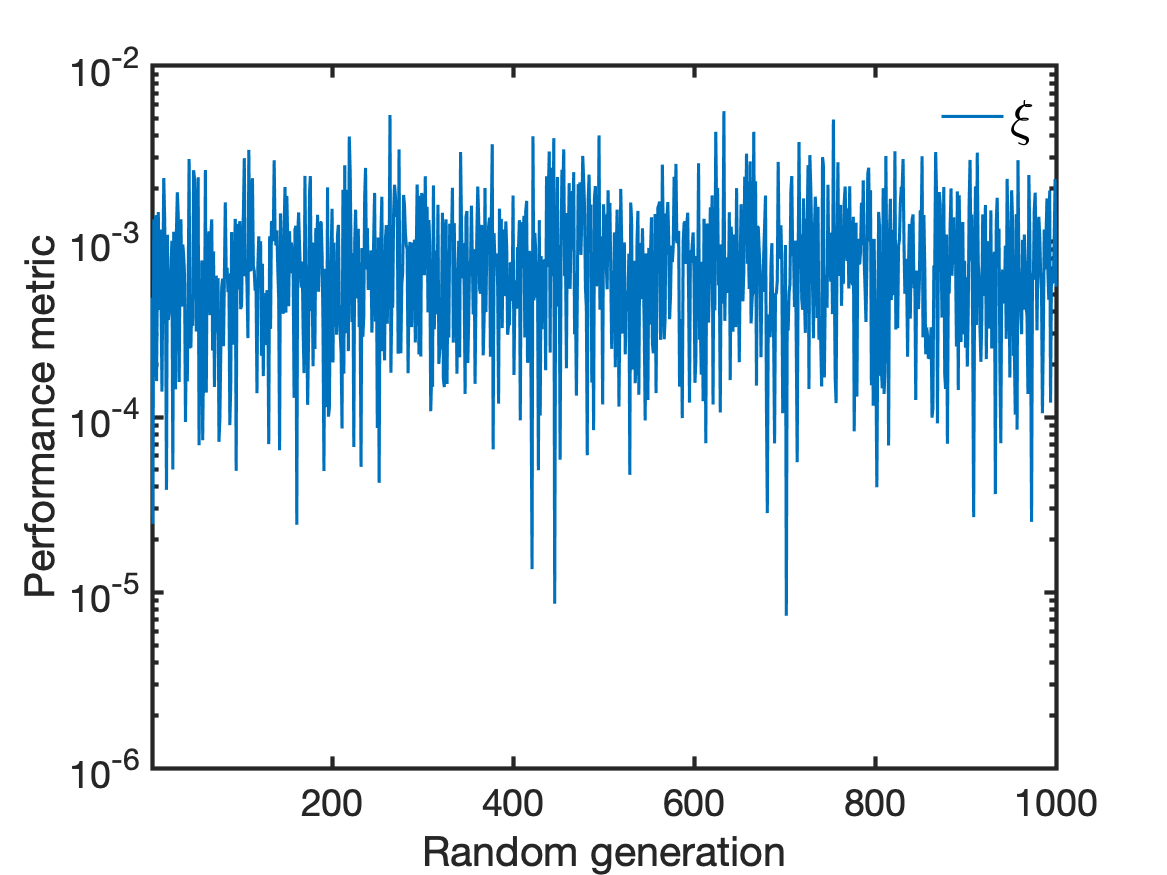}
\caption{Performance metric as defined in \eq{criterion} for 1000 random generations. Three generations show a very low $\xi$ value which indicates a closeness to the target generation. However, the most likely values are different from the minimum values.}
\label{fig:performance}
\end{figure}

\begin{figure}[t]
\centering
\includegraphics[width=0.32\textwidth]{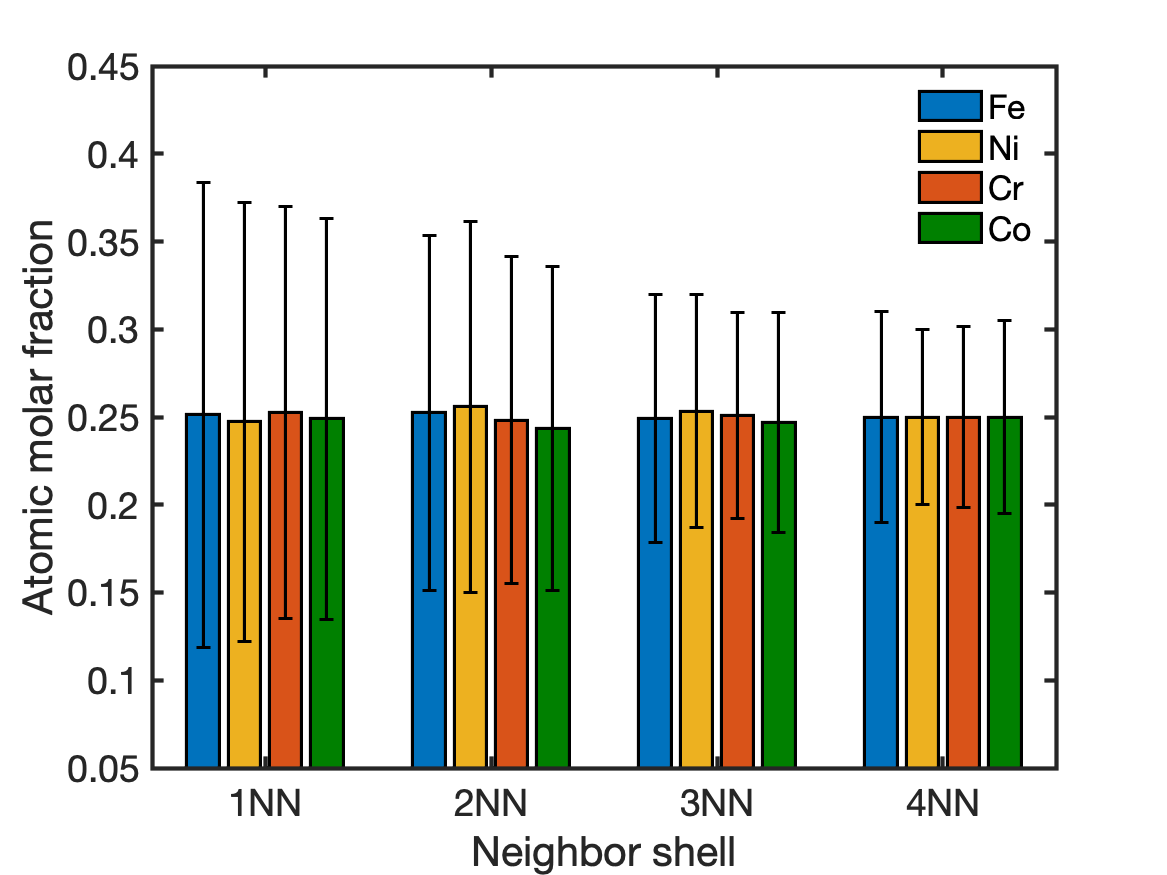}
\includegraphics[width=0.32\textwidth]{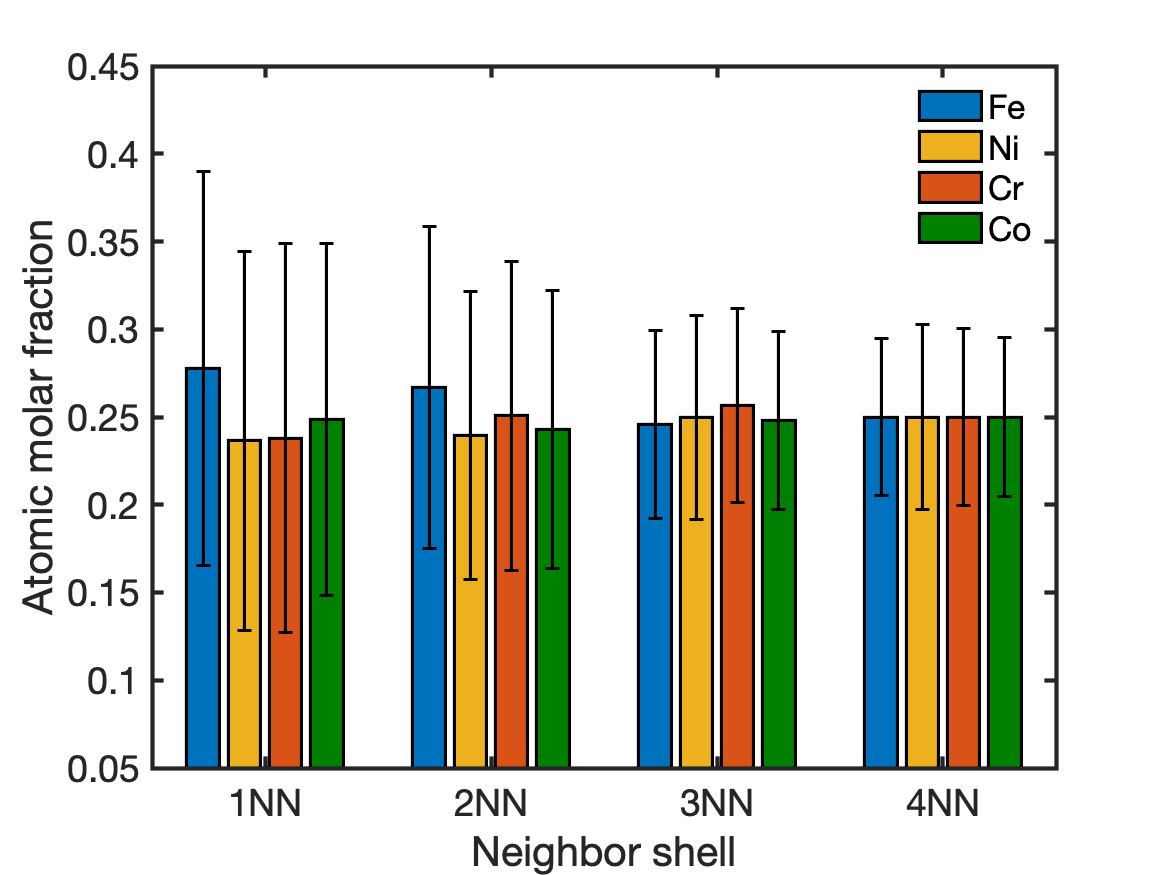}
\includegraphics[width=0.32\textwidth]{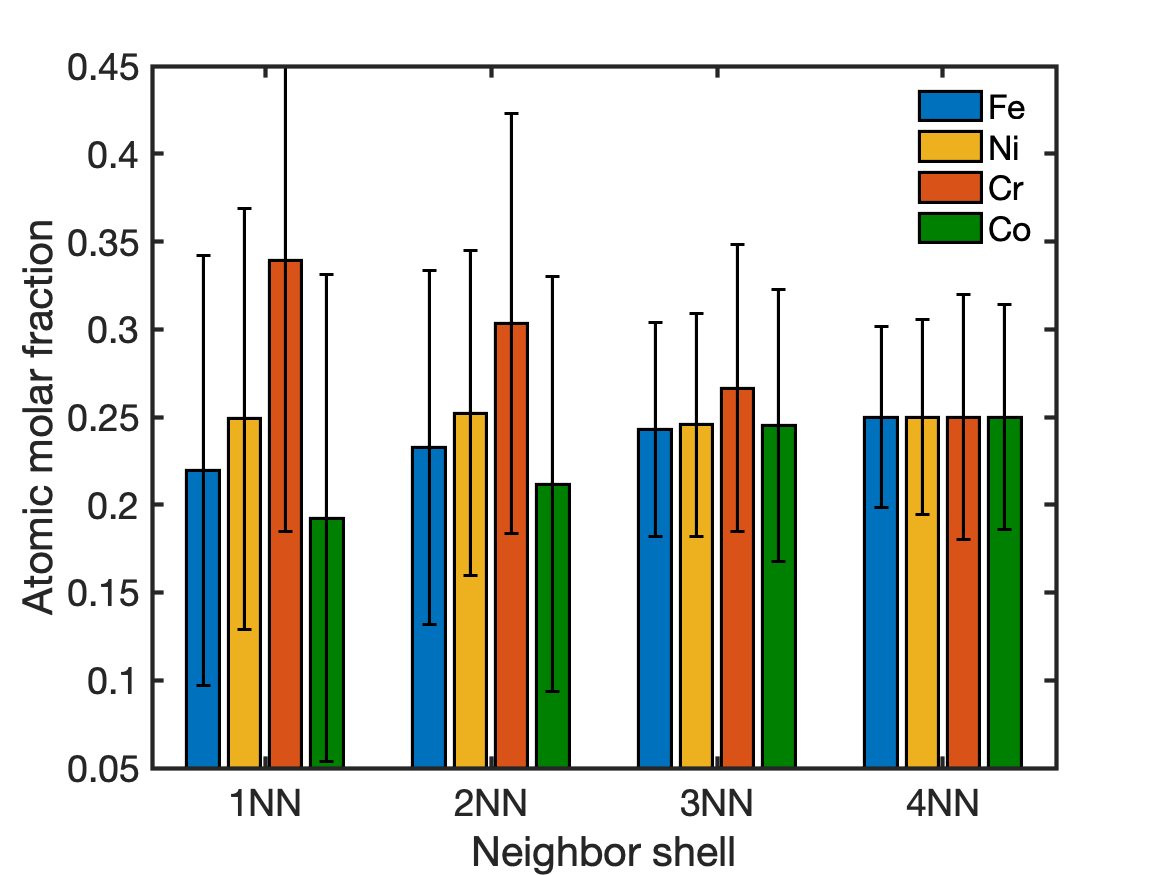}
\caption{Average chemical environment for an equiatomic HEA random generation as a function of the neighbor shells. The standard deviation for each element and each shell is also shown.}
\label{fig:chem_env}
\end{figure}

\newpage
\section{Vacancy formation energy in an High Entropy Alloy}

The vacancy formation energy is given by \cite{PhysRevMaterials.2.013602}
\begin{equation} \label{Eq:vfe}
    f^v=E_d(N_s-1)-E_0(N_s)+\mu_d \,\,,
\end{equation}
where $f^v$ is the vacancy formation energy, $E_d(N_s-1)$ is the energy of the system with the vacancy, $E_0(N_s)$ is the energy of the perfect system (i.e., without any defects), and $\mu_d$ is the chemical potential of the defect species. 
For a four component CoFeCrNi alloy, the chemical potentials can be calculated using the relations
\begin{eqnarray}
\mu_\mathrm{Co}-\mu_\mathrm{Cr} = E^{\mathrm{Co} \rightarrow \mathrm{Cr}} - E_0 =: A_\mathrm{CoCr} \,\,, \\
\mu_\mathrm{Co}-\mu_\mathrm{Fe} = E^{\mathrm{Co} \rightarrow \mathrm{Fe}} - E_0 =: A_\mathrm{CoFe} \,\,, \\
\mu_\mathrm{Co}-\mu_\mathrm{Ni} = E^{\mathrm{Co} \rightarrow \mathrm{Ni}} - E_0 =: A_\mathrm{CoNi} \,\,, \\
\mu_\mathrm{Cr}-\mu_\mathrm{Fe} = E^{\mathrm{Cr} \rightarrow \mathrm{Fe}} - E_0 =: A_\mathrm{CrFe} \,\,, \\
\mu_\mathrm{Cr}-\mu_\mathrm{Ni} = E^{\mathrm{Cr} \rightarrow \mathrm{Ni}} - E_0 =: A_\mathrm{CrNi} \,\,, \\
\mu_\mathrm{Fe}-\mu_\mathrm{Ni} = E^{\mathrm{Fe} \rightarrow \mathrm{Ni}} - E_0 =: A_\mathrm{FeNi} \,\,. \\
\end{eqnarray}
Subject to the constraints:
\begin{equation}
    N_\mathrm{Co}\mu_\mathrm{Co} + N_\mathrm{Cr}\mu_\mathrm{Cr} + N_\mathrm{Fe}\mu_\mathrm{Fe} + N_\mathrm{Ni}\mu_\mathrm{Ni} = E_0 \,,
\end{equation}
and 
\begin{equation}
    N_\mathrm{Co} + N_\mathrm{Cr} + N_\mathrm{Fe} + N_\mathrm{Ni} = N_s,
\end{equation}
where $N_\mathrm{Co}$, $N_\mathrm{Cr}$, $N_\mathrm{Fe}$ and $N_\mathrm{Ni}$ are the number of Co, Cr, Fe and Ni atoms in the simulation cell, respectively. $N_s$ is the total number of atoms on the system.
For the equiatomic alloy ($N_\mathrm{Co} = N_\mathrm{Cr} = N_\mathrm{Fe} = N_\mathrm{Ni} = \frac{N_s}{4}$), the above equations can be used to write
\begin{eqnarray}\label{Eq:ChemPotFinal}
\mu_\mathrm{Co} =\frac{1}{4} (\frac{4E_{0}}{N_s} + A_\mathrm{CoCr} + A_\mathrm{CoFe} + A_\mathrm{CoNi}) \,\,, \\
\mu_\mathrm{Cr} = \mu_\mathrm{Co} - A_\mathrm{CoCr} \,\,, \\
\mu_\mathrm{Fe} = \mu_\mathrm{Co} - A_\mathrm{CoFe} \,\,, \\
\mu_\mathrm{Ni} = \mu_\mathrm{Co} - A_\mathrm{CoNi} . 
\end{eqnarray}
Once, the chemical potentials are determined using the above Eq.~\ref{Eq:ChemPotFinal}, we use Eq.~\ref{Eq:vfe} to calculate the vacancy formation energies. The values of the chemical potentials using DFT calculations were $\mu_\mathrm{Co}$ = -14.403 eV$\cdot$atom$^{-1}$, $\mu_\mathrm{Cr}$ = -8.904 eV$\cdot$atom$^{-1}$, $\mu_\mathrm{Fe}$ = -12.579 eV$\cdot$atom$^{-1}$, and $\mu_\mathrm{Ni}$ = -16.385 eV$\cdot$atom$^{-1}$. For classical molecular static simulations, the chemical potentials were $\mu_\mathrm{Co} = -4.29~\pm~0.03$ eV$\cdot$atom$^{-1}$, $\mu_\mathrm{Cr} = -4.571~\pm~0.097$ eV$\cdot$atom$^{-1}$, $\mu_\mathrm{Fe} = -4.329~\pm~0.05$ eV$\cdot$atom$^{-1}$, and $\mu_\mathrm{Ni} = -4.255~\pm~0.03$ eV$\cdot$atom$^{-1}$

\section{Atomic vibrational frequencies}
Vibrational frequencies for atoms in the equimolar CoFeCrNi MPEAs has been computed using the quasi-harmonic approximation of the hessian, computed using the thermalized interatomic potential at $T=300$ K. Details of calculations can be found in other works \cite{Ariza:2012,Ponga2017a,Ponga2018c,Mendez:2021}. 

\begin{figure}[!htb]
\centering
\includegraphics[width=0.45\textwidth]{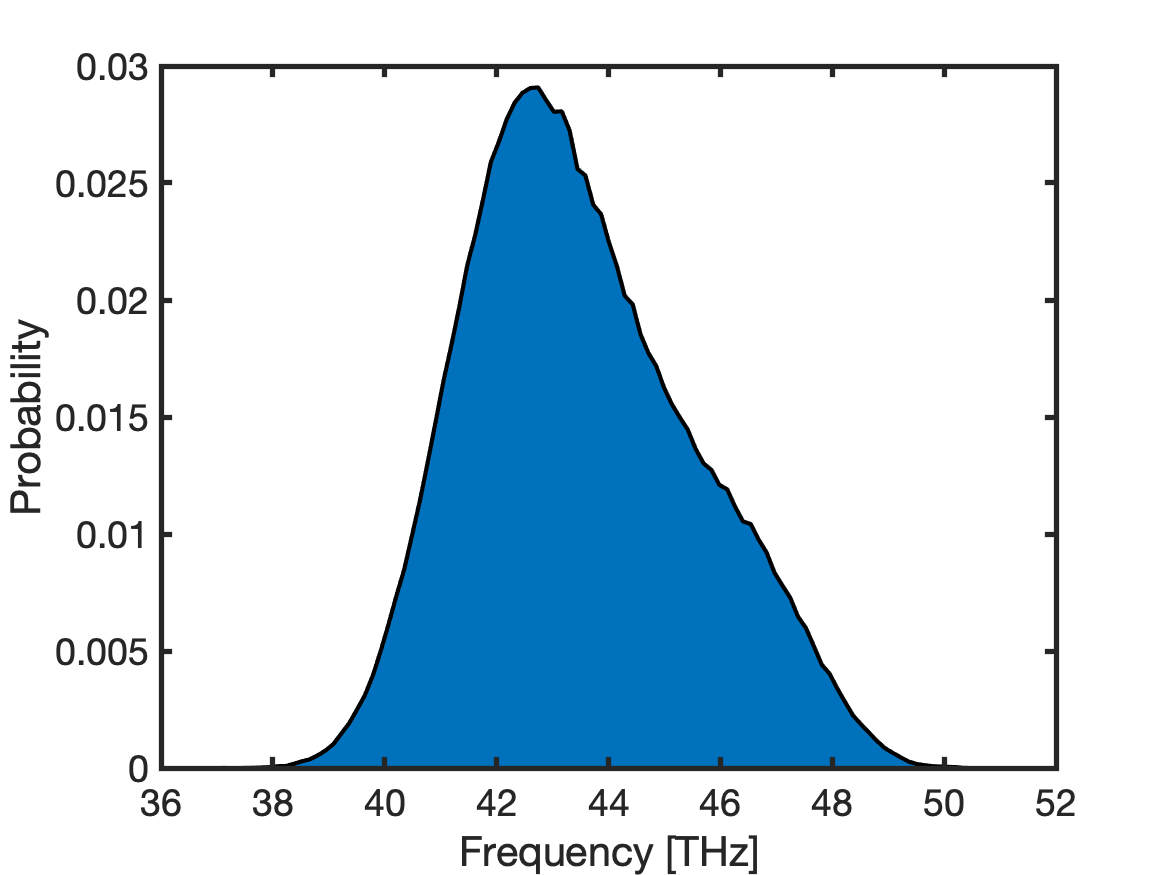}
\caption{Distribution of the vibrational frequencies of atoms for an equimolar CoFeCrNi MPEAs.}
\label{fig:VibFreq}
\end{figure}

\newpage
\section{Machine learning predictions of self-interstitial atoms formation energies}
\begin{figure}[!htb]
\centering
\includegraphics[width=0.45\textwidth]{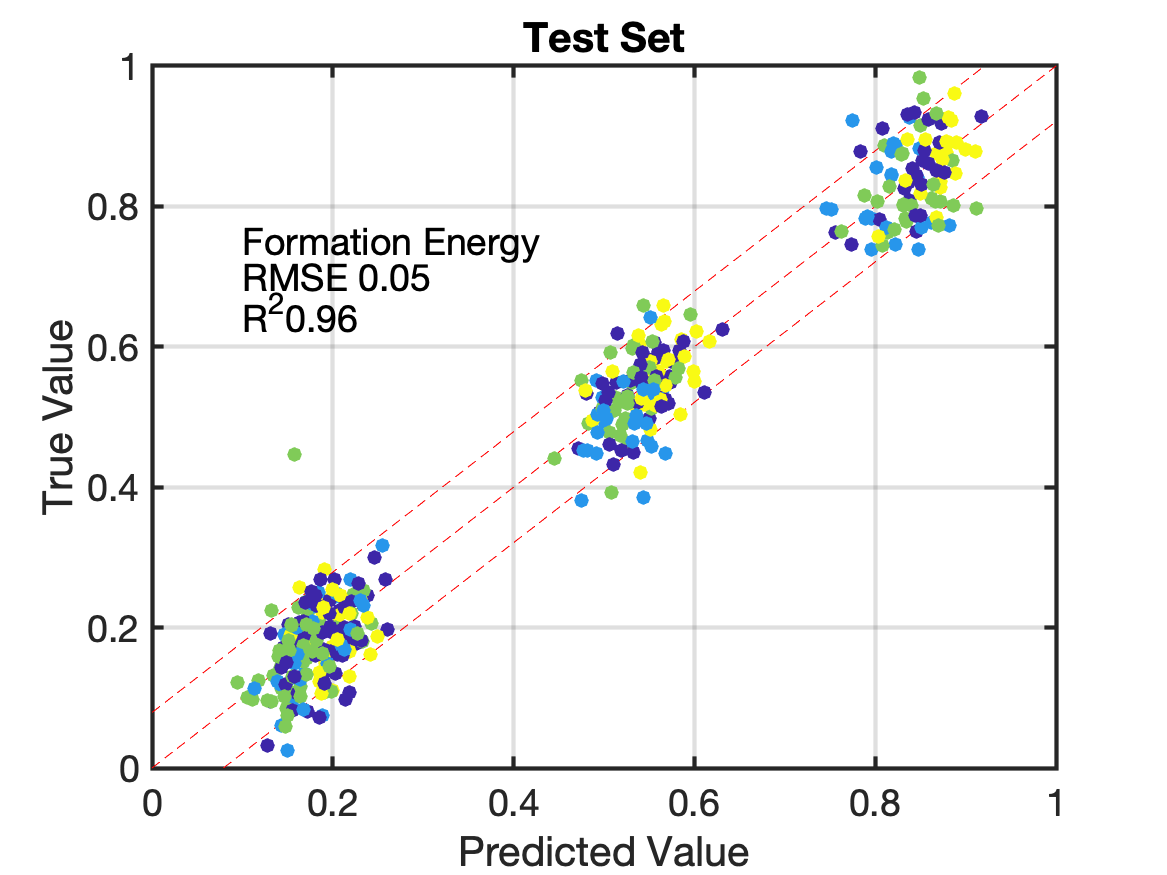}
\caption{Predicted \emph{vs.} true values of self interstitial atom energies for a CoFeCrNi high-entropy alloy as predicted with the trained CNN using the test set.}
\label{fig:SIAAll}
\end{figure}

\newpage

\section{Probability distribution functions for all vacancies as function of temperature.}
\begin{figure}[!htb]
\centering
\includegraphics[width=0.45\textwidth]{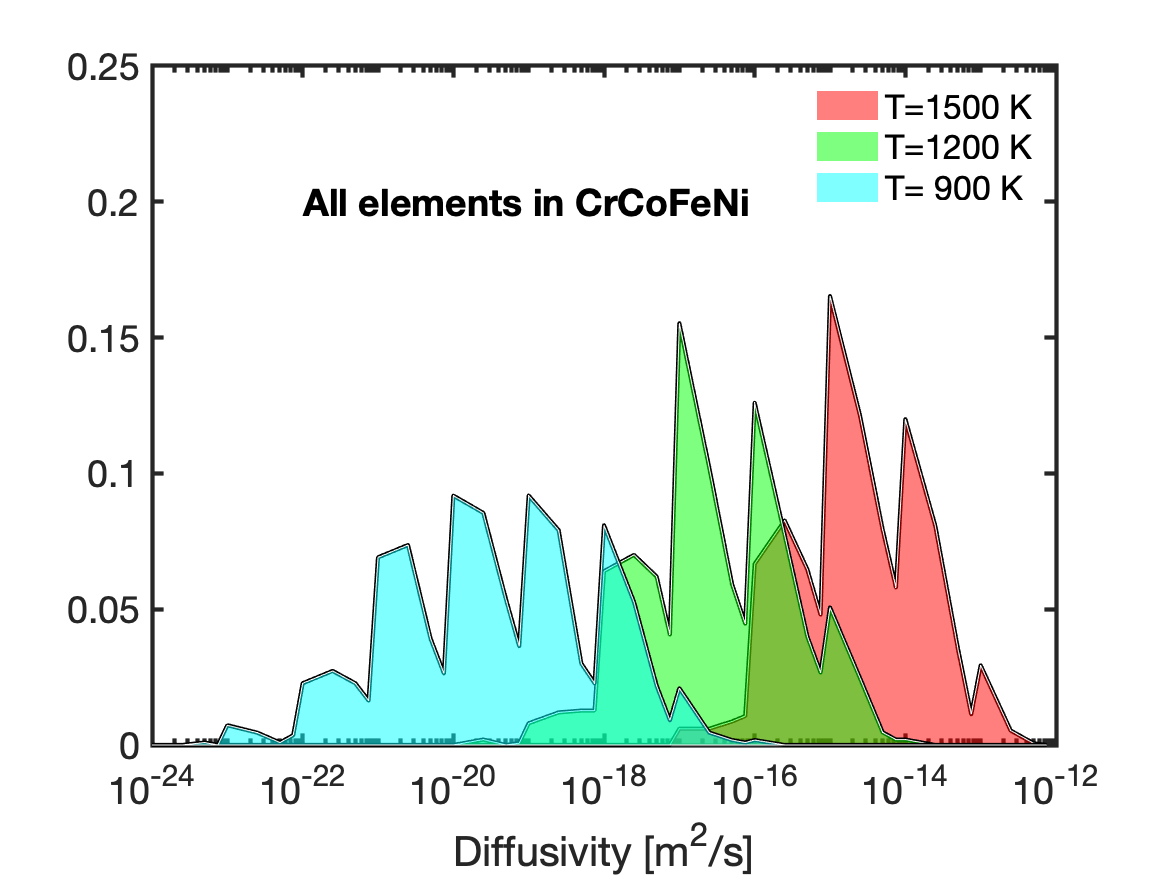}
\caption{Probability distribution for vacancy diffusivities for all elements in an equimolar CoFeCrNi MPEA computed with the MLMC method for a temperature range between $T=900 - 1500$~K. }
\label{fig:distTemp}
\end{figure}

\end{document}